\documentclass[aps,prl,amsmath,amssymb,floatfix,twocolumn,groupedaddress,10pt]{revtex4-2}
\usepackage{times}
\usepackage{bbold}
\usepackage{bm}
\usepackage{graphicx} 
\usepackage{color}
\usepackage{hyperref}
\usepackage{textcomp}
\usepackage{amssymb} 
\usepackage{wasysym}
\usepackage{epstopdf}

\usepackage{dcolumn} 
  
\usepackage{multibib} 
\newcites{supp}{Supplemental References}

\newcommand{\Moire}{Moir{\' e} }
\newcommand{\moire}{moir{\' e} }

\begin{document}

\title{General scatterings and electronic states in the quantum-wire network of \moire systems}

\author{Chen-Hsuan Hsu$^{1,2,3}$} 
\author{Daniel Loss$^{3,4}$}
\author{Jelena Klinovaja$^{4}$}

\affiliation{$^{1}$Yukawa Institute for Theoretical Physics, Kyoto University, Kyoto 606-8502, Japan}
\affiliation{$^{2}$Institute of Physics, Academia Sinica, Taipei 115, Taiwan}
\affiliation{$^{3}$RIKEN Center for Emergent Matter Science, Wako, Saitama 351-0198, Japan}
\affiliation{$^{4}$Department of Physics, University of Basel, Klingelbergstrasse 82, CH-4056 Basel, Switzerland}

\date{\today}

\begin{abstract}
We investigate electronic states in a two-dimensional network consisting of interacting quantum wires, a model adopted for twisted bilayer systems. We construct general operators which describe various scattering processes in the system. In a twisted bilayer structure, the \moire periodicity allows for generalized umklapp scatterings, leading to a class of correlated states at certain fractional fillings.  
We identify scattering processes which can lead to an insulating gapped bulk with gapless chiral edge modes at fractional fillings, resembling the quantum anomalous Hall effect recently observed in twisted bilayer graphene. Finally, we demonstrate that the description can be useful in predicting spectroscopic and transport features to detect and characterize the chiral edge modes in the moir{\'e}-induced correlated states.
\end{abstract}

\maketitle

\Moire bilayer structures provide a platform for strongly correlated systems, where unconventional states of matter emerge~\cite{Andrei:2020,Balents:2020,Andrei:2021,Kennes:2021} 
as a consequence of flat energy bands~\cite{Bistritzer:2011}.  
Since the discovery of correlated insulating states and superconductivity in twisted bilayer graphene (TBG)~\cite{Cao:2018a,Cao:2018b}, various exotic states or features have been observed~\footnote{We additionally note an earlier experimental study on the electronic structure of \moire MoS$_2$/WSe$_2$ heterobilayers~\cite{Zhang:2017}.}, including nematicity~\cite{Choi:2019,Jiang:2019,Kerelsky:2019,Cao:2020}, pressure-enhanced superconductivity~\cite{Yankowitz:2019}, 
strange metal~\cite{Polshyn:2019,Cao:2020b}, cascade of transitions~\cite{Wong:2020,Zondiner:2020}, orbital magnetism~\cite{Lu:2019,Sharpe:2021},
independent superconducting and correlated insulating states~\cite{Arora:2020,Saito:2020,Stepanov:2020},
fragile correlated states against twist angle disorder~\cite{Uri:2020}, entropy-driven phase transition~\cite{Saito:2021a,Rozen:2021}, unconventional superconductivity~\cite{Oh:2021}, and spin-orbit-driven ferromagnetism~\cite{Lin:2022}.

In addition to features that resemble existing strongly correlated systems such as cuprates and iron-based superconductors, there are observations suggesting the existence of topological phases. Specifically, nonlocal transport demonstrated the presence of chiral edge modes at 3/4 filling in TBG~\cite{Sharpe:2019}, accompanied by the quantization of Hall resistance at zero magnetic fields~\cite{Serlin:2020}. 
More recent studies revealed a series of quantum anomalous Hall or Chern insulators with Chern numbers $C = \pm 1, \pm 2$ and $\pm 3$ at $\pm 3/4, \pm1/2$ and $\pm 1/4$ fillings, respectively~\cite{Nuckolls:2020,Choi:2021,Das:2021}.
Furthermore, there is experimental indication of a many-body origin for the topological phases~\cite{Nuckolls:2020,Choi:2021,Das:2021,Saito:2021b,Stepanov:2021,Park:2021b,Lin:2022,Tseng:2022}. 
The observations on various electronic states motivated theoretical studies on TBG~\cite{Kennes:2018,Koshino:2018,Po:2018,Wu:2018,Kang:2019,Lian:2019,Seo:2019,Bultinck:2020a,Bultinck:2020b,Christos:2020,Xie:2020,Hejazi:2021,Khalaf:2021,Liu:2021,Parker:2021,Shavit:2021,Wagner:2022} 
and the development of \moire electronics, including structures beyond bilayers~\cite{Burg:2019,Chen:2019,Chen:2019a,Cao:2020c,Cao:2021,Liu:2020,Shen:2020,Hao:2021,Liu:2021b,Park:2021a} and materials other than graphene~\cite{Regan:2020,Shimazaki:2020,Tang:2020,Wang:2020,Xu:2020,Zhang:2020}.
 
\begin{figure}[t]
\includegraphics[width=\linewidth]{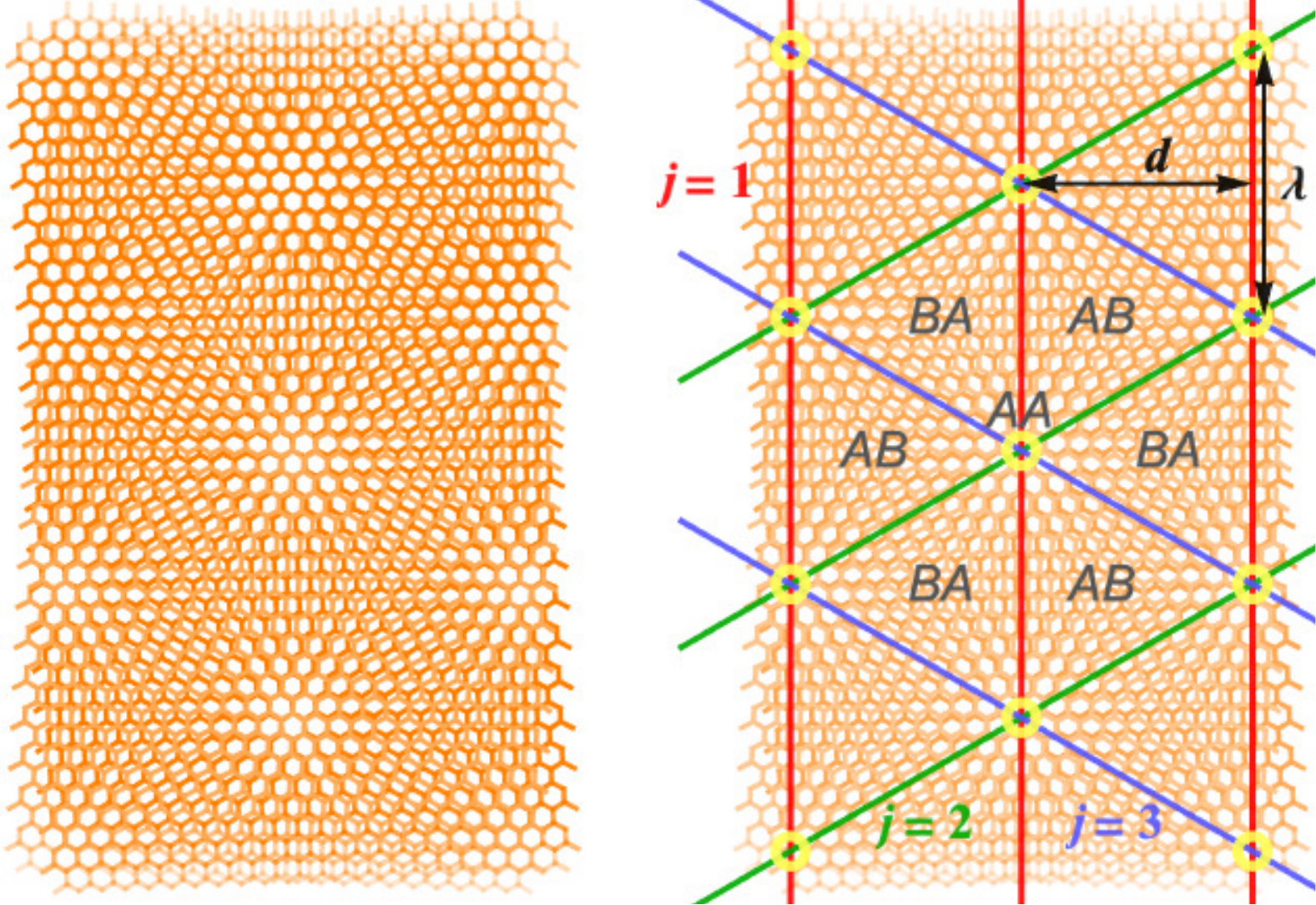} 
\caption{\Moire pattern and quantum-wire network of the TBG. 
When two graphene monolayers (orange) are stacked with a misalignment, there appears a \moire pattern with the wavelength $\lambda = a_0/[2 \sin (\theta/2)] $, monolayer lattice constant $a_0$, and the angle $\theta$ between the layers. 
The \moire pattern results in three sets of parallel quantum wires, plotted in distinct colors and labeled by $j$, with the interwire distance $d = \sqrt{3} \lambda /2 $.
}
\label{Fig:Lattice}
\end{figure}

A major theoretical challenge in strongly correlated \moire systems involves incorporating many-body effects with numerous atoms due to the large \moire unit cell. It is thus crucial to identify the relevant degrees of freedom to construct an effective model for efficient quantitative analysis.
Remarkably, correlated phenomena in TBG can be investigated in the context of (Tomonaga-)Luttinger liquids, which inherently includes electron-electron interactions~\cite{Tomonaga:1950,Luttinger:1963,Haldane:1981}. 
Specifically, in the presence of an interlayer potential difference, one-dimensional channels emerge at domain walls between AB- and BA-stacking regions~\cite{San-Jose:2013,Nam:2017,Efimkin:2018} and form a triangular quantum-wire network illustrated in Fig.~\ref{Fig:Lattice}; we also note spectroscopic~\cite{Huang:2018,Choi:2019,Kerelsky:2019,Jiang:2019,Xie:2019} 
and transport~\cite{Rickhaus:2018} features of the domain-wall network~\footnote{The STM experiments revealed additional features around AA-stacking regions, which inspired Refs.~\cite{Song:2022,Shi:2022,Lau:2023} to construct a heavy-fermion model.
}.
These findings motivated theoretical studies on network models based on Luttinger liquids~\cite{Wu:2019,Chou:2019,Chen:2020,Chou:2021,Lee:2021}, reminiscent of earlier works on (crossed) sliding Luttinger liquids proposed for cuprates~\cite{Emery:2000,Vishwanath:2001,Mukhopadhyay:2001a,Mukhopadhyay:2001b} and the coupled-wire constructions of various quantum Hall states~\cite{Kane:2002,Klinovaja:2013b,Klinovaja:2014,Klinovaja:2014b,Neupert:2014,Sagi:2014,Teo:2014,Klinovaja:2015,Santos:2015,Imamura:2019,Meng:2020}. 
From a different perspective, \moire systems provide mesoscopic realizations of coupled-wire systems originally proposed for entirely distinct systems~\cite{Emery:2000,Vishwanath:2001,Mukhopadhyay:2001a,Mukhopadhyay:2001b}.

In this work, we extend the network models~\cite{Wu:2019,Chou:2019,Chen:2020,Chou:2021,Lee:2021} to explore the possibility for topological phases in \moire systems.
We construct operators describing general scattering processes based on conservation laws and investigate the resulting electronic states. 
In \moire structures, the periodic potential allows for generalized umklapp scatterings, which lead to correlated states at fractional fillings. 
Remarkably, we identify processes that lead to a gapped bulk with gapless modes along the edges, resembling the observed Chern insulators in TBG~\cite{Sharpe:2019,Serlin:2020,Nuckolls:2020,Choi:2021,Das:2021,Saito:2021b,Stepanov:2021,Park:2021b}. 
Furthermore, we demonstrate that this description can be useful by making concrete predictions for spectroscopic and transport features. In addition to TBG, our mechanism can apply to 
other nanoscale systems forming arrays of one-dimensional channels, such as twisted \moire bilayers formed by WTe$_{2}$~\cite{Wang:2022} or topological insulators~\cite{Fujimoto:2022,Tateishi:2022}, as well as strain-engineered graphene~\cite{Hsu:2020n}. 

\begin{figure}[t]
\centering
\includegraphics[width=0.46\linewidth]{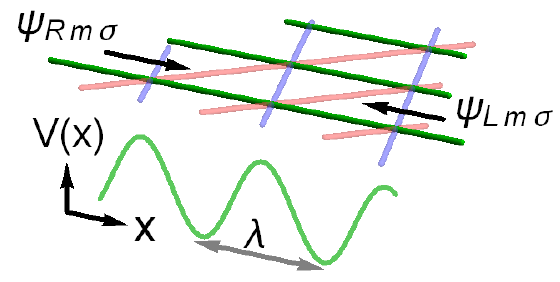}
\hspace{0.0\linewidth}
\includegraphics[width=0.52\linewidth]{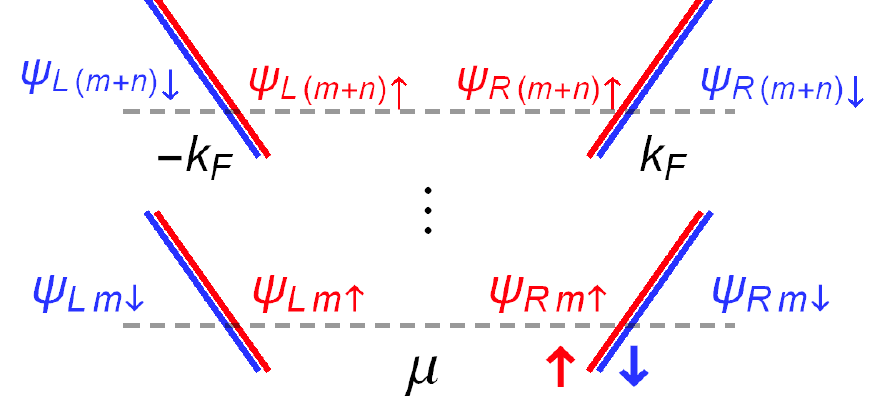}
\caption{Quantum-wire network in a \moire structure.
Left: For each wire, we define the local coordinate $x$ and fermion fields $\psi_{\ell m\sigma}$, which experience a periodic potential $V(x)$ generated by the \moire structure.
Right: Each array consists of parallel wires with the chemical potential $\mu$ and Fermi wave vector $k_F$, where we linearize the energy dispersion and bosonize the fields with Eq.~\eqref{Eq:bosonization}.
}
\label{Fig:wires}
\end{figure}

{\it Bosonization.} We introduce the fermion field $\psi_{\ell m\sigma}^{(j)} $ with the array index $j \in \{ 1,2,3\}$, wire index $m \in [1,N_{\perp}]$ within each array, the index $\ell \in \{ R \equiv  +, L \equiv - \}$ labeling the moving direction, and spin $\sigma \in \{\uparrow \, \equiv + , \downarrow \, \equiv - \}$; see Fig.~\ref{Fig:wires}. 
The fermion field can be bosonized as
\begin{eqnarray}
\psi_{\ell m\sigma}^{(j)} (x) &=&  \frac{ U_{\ell m \sigma}^{j}}{\sqrt{2\pi a}} e^{i \ell k_{F}x}   \nonumber \\
&& \times e^{\frac{-i}{\sqrt{2}}[\ell \phi_{cm}^{j} (x) - \theta_{cm}^{j} (x) + \ell \sigma \phi_{sm}^{j} (x) - \sigma \theta_{sm}^{j} (x)]}, 
\label{Eq:bosonization}
\end{eqnarray}
with the Klein factor $U_{\ell m \sigma}^{j}$, short-distance cutoff $a$, local coordinate $x$, Fermi wave vector $k_F$ (identical for all wires), and the index 
$\xi \in \{c \equiv + , s \equiv - \}$ 
for the charge/spin sector of the boson fields $\phi_{\xi m}^{j}$ and $ \theta_{\xi m}^{j}$, satisfying
\begin{eqnarray} 
\left[ \phi_{\xi m}^{j}(x), \theta_{\xi' m'}^{j'}(x' )\right] &=& i \frac{\pi}{2} {\rm sign}(x'-x) \delta_{j j'} \delta_{\xi \xi'} \delta_{m m'}.
\end{eqnarray}
Below we omit the Klein factor and $x$ whenever possible.

The unperturbed Hamiltonian $H_0 + H_{\rm fs}$ describes a crossed sliding Luttinger liquid at the fixed point~\cite{Chen:2020}, with the kinetic energy $H_0$ and marginally relevant forward scattering terms $H_{\rm fs}$ quadratic in the density operator $\propto \partial_x \phi_{cm}^{j} $.
In addition, there exist intrawire or interwire backscattering processes, arising from electron-electron interactions and/or tunnelings, which can destabilize the fixed point characterized by the quadratic terms, as those in coupled-wire systems~\cite{Kane:2002,Klinovaja:2013b,Klinovaja:2014,Klinovaja:2014b,Neupert:2014,Sagi:2014,Teo:2014,Klinovaja:2015,Santos:2015,Imamura:2019,Meng:2020}. Since the bandwidth $W$ serves as high-energy cutoff~\cite{Giamarchi:2003}, the dimensionless coupling $g/W$, with the strength $g$ characterizing a general scattering, takes a larger value in (quasi-)flat-band systems, allowing for higher-order scatterings to play a more significant role. 
As in Refs.~\cite{Kane:2002,Teo:2014}, we do not specify $H_{\rm fs}$; for demonstration, a specific model~\cite{Emery:2000,Vishwanath:2001,Mukhopadhyay:2001a,Mukhopadhyay:2001b,Chen:2020} is presented in Supplemental Material (SM)~\cite{SM}.
Below we construct operators describing general scatterings, including higher-order processes (previously discussed in multiband wires~\cite{Shavit:2019,Hsu:2020}), and discuss the resulting electronic states. 

{\it General scattering operator.}
We consider the operator, 
\begin{eqnarray}
O_{\{s_{\ell p\sigma}^j \}} &=& \sum_{m=1}^{N_{\perp}} \prod_{p=0} \prod_{j=1}^{3}  \big[\psi_{R (m+p) \uparrow}^{(j)} \big]^{s_{Rp\uparrow}^j}  \big[ \psi_{L (m+p) \uparrow}^{(j)} \big]^{s_{Lp\uparrow}^j} \nonumber \\
&& \hspace{24pt} \times  \big[\psi_{R (m+p) \downarrow}^{(j)}\big]^{s_{Rp\downarrow}^j} \big[ \psi_{L (m+p) \downarrow}^{(j)} \big]^{s_{Lp\downarrow}^j} ,  
\label{Eq:O_gs}
\end{eqnarray}
where the subscript $\{ s_{\ell p\sigma}^j \}$ denotes an integer set for all values of $(j, \ell, p , \sigma)$ with $p \in $ integers. The set characterizes $O$; a negative value implies Hermitian conjugate: $\psi^{s} \equiv ( \psi^{\dagger} )^{| s |}$ for $s < 0$. 
A nonzero $s$ for a given $p$ indicates that the $p$-th nearest neighbor wires participate in the scattering.
While $O$ can in principle involve any number of wires, physically one expects $s$ to vanish for large $p$ in systems subject to finite-range interactions. 

The operator $O$ describes scatterings within an array when $s$ is nonzero for a single $j$ value. The corresponding renormalization-group (RG) relevance condition is given by $\Delta_{{s_{\ell p\sigma}^{j}}} <2$, where the scaling dimension $\Delta_{{s_{\ell p\sigma}^{j}}}$ is determined by $H_0 + H_{\rm fs}$.
In a network consisting of crossed wires, interarray scatterings can occur at wire intersections~\cite{Mukhopadhyay:2001a,Mukhopadhyay:2001b,Chou:2019,Chen:2020}, as characterized by Eq.~\eqref{Eq:O_gs} with nonzero $s$ for multiple $j$ values. Refs.~\cite{Chou:2019,Chen:2020} showed that such scatterings can induce superconducting and insulating phases in \moire bilayers. However, the RG relevance condition in this case is more stringent: $\Delta_{{s_{\ell p\sigma}^{j}}} <1$, since the corresponding operator enters the effective action without involving the spatial integral~\cite{SM}.
Furthermore, the interarray scatterings are independent of the filling factor. To explore correlated states from more RG relevant scatterings, below we examine scatterings within an array and suppress $j$.

We start with the constraints on possible $s_{\ell p\sigma}$ values. 
In the absence of proximity-induced ``external'' pairing, the global particle number or charge is conserved, giving
\begin{eqnarray}
\sum_{p,\sigma} (s_{Rp\sigma}+s_{Lp\sigma}) = 0.
\label{Eq:ChargeConser}
\end{eqnarray}
For clean systems, the momentum conservation gives additional constraint. Here, the \moire structure plays an important role, as it creates a periodic potential, which partially relaxes the constraint from the momentum conservation. 
As illustrated in Fig.~\ref{Fig:wires}, electrons experience a \moire potential with a spatial period of $\lambda$.
This leads to a generalized condition for momentum conservation,
\begin{eqnarray}
k_{F} \sum_{p,\sigma}  (s_{Rp\sigma} - s_{Lp\sigma}) = \frac{2\pi}{\lambda} \times ~{\rm integer},
\label{Eq:MomentumConser}
\end{eqnarray}
which allows us to organize $O_{\{s_{\ell p\sigma} \}} $ into two categories. 

In the first category, scatterings are allowed for any $k_{F}$ independent of the filling factor, provided that the coefficients satisfy
\begin{eqnarray}
 \sum_{p,\sigma}  (s_{Rp\sigma} - s_{Lp\sigma}) = 0.
\label{Eq:conditionLR}
\end{eqnarray}
Together with the constraint in Eq.~\eqref{Eq:ChargeConser}, we get 
\begin{eqnarray}
 \sum_{p,\sigma} s_{Rp\sigma} =  \sum_{p,\sigma} s_{Lp\sigma} = 0,
 \label{Eq:conditionLR2}
\end{eqnarray}
meaning that the numbers of the left- and right-moving particles are individually conserved.  
We refer to these processes as {\it conventional scatterings}, which characterize electronic states corresponding the ``crystalline states'' in Ref.~\cite{Kane:2002}. 

At certain fillings, on the other hand, another category of scatterings can take place even when Eq.~\eqref{Eq:conditionLR} is not fulfilled. 
The momentum difference due to the number imbalance between the left- and right-moving particles can be compensated by the ``crystal momentum'' proportional to the reciprocal lattice vector $2\pi/\lambda$.
With Eqs.~\eqref{Eq:ChargeConser}--\eqref{Eq:MomentumConser} and the relation between the filling factor and Fermi wave vector $\nu = k_F \lambda / \pi$~\cite{Giamarchi:2003}, we get a condition on the filling factor,
\begin{eqnarray}
\nu &=& \frac{P}{  \sum_{p,\sigma}  s_{Rp\sigma}  } ,
\label{Eq:nu_condition}
\end{eqnarray}
with a nonzero integer $P$.
In our description, $\nu = 1$ corresponds to 4 electrons per \moire unit cell in TBG~\cite{Chou:2019,Chen:2020}. 
Since these processes are feasible owing to the presence of the \moire periodic potential, in analogy to Refs.~\cite{Giamarchi:1991,Giamarchi:2003}, we refer to the second category as {\it \moire umklapp scatterings} and the corresponding states of matter {\it \moire correlated states}.

For both categories, the bosonization in Eq.~\eqref{Eq:bosonization} gives
\begin{eqnarray}
O_{\{s_{\ell p\sigma}\}} &=& \sum_{m=1} {\rm Exp} \Big\{ \frac{i}{\sqrt{2}}
\sum_{p} \big[ S_{p,c} \phi_{c (m+p)} + \bar{S}_{p,c} \theta_{c (m+p)} \nonumber \\
&& \hspace{42pt}  + S_{p,s} \phi_{s (m+p)} + \bar{S}_{p,s} \theta_{s (m+p)} \big] \Big\}, 
\label{Eq:O_general}
\end{eqnarray} 
with the coefficients,
\begin{subequations}
\label{Eq:S_coefficient}
\begin{eqnarray}
S_{p,\xi} & = & s_{Lp\uparrow} - s_{Rp\uparrow}  +\xi ( s_{Lp\downarrow} - s_{Rp\downarrow}), \\
 \bar{S}_{p,\xi} &=& s_{Lp\uparrow} + s_{Rp\uparrow}  + \xi ( s_{Lp\downarrow} + s_{Rp\downarrow}).
\end{eqnarray}
\end{subequations}
The global charge conservation requires $\sum_p \bar{S}_{p,c} = 0$. The momentum conservation requires $\sum_p S_{p,c} = 0$ for conventional scatterings and $\nu \sum_p S_{p,c} =2P $ for \moire umklapp scatterings. 
If the charge (spin) is conserved for a fixed $p$, the coefficient $ \bar{S}_{p,c}$ ($\bar{S}_{p,s}$) vanishes. 
While there is in general no constraint on $\bar{S}_{p,s}$, for simplicity we choose $\bar{S}_{p,s}=0$, 
as operators with nonzero $\bar{S}_{p,s}$ are typically less RG relevant.  

The conventional scatterings fulfilling Eq.~\eqref{Eq:conditionLR2} include charge-density-wave couplings, Josephson couplings, and hoppings. They lead to charge density wave, superconducting, and Fermi liquid states, respectively~\cite{SM}. In addition, the twisted structure enables \moire umklapp scatterings, which we discuss below.

\begin{figure}[t]
\centering
 \includegraphics[width=0.48\linewidth]{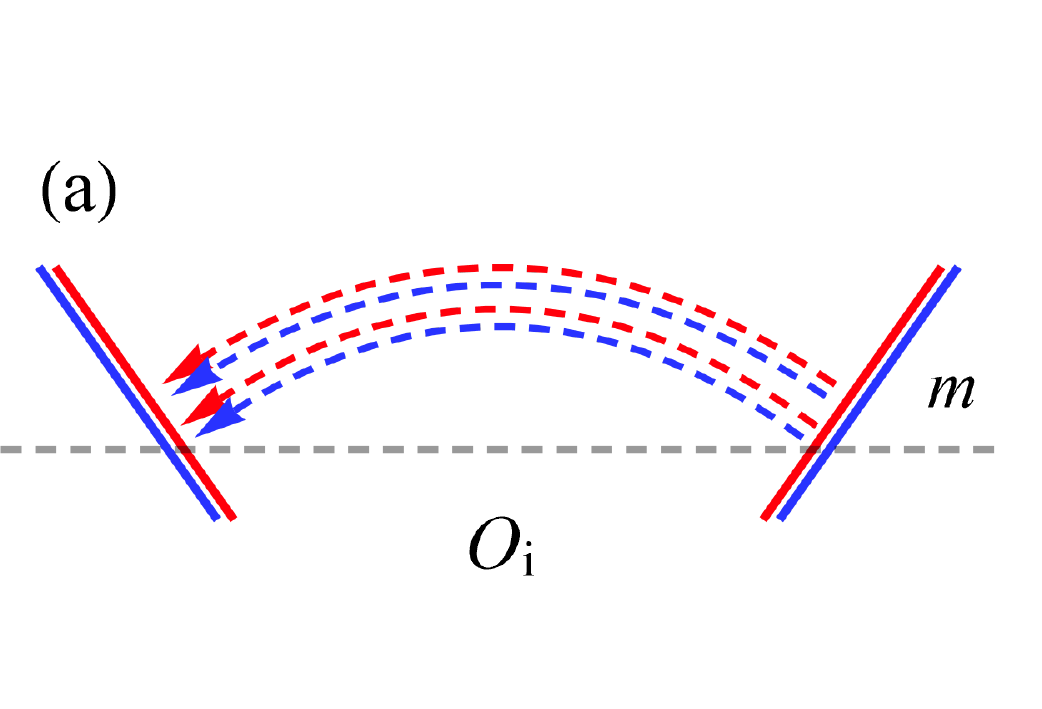}
 \hspace{0.01\linewidth}
 \includegraphics[width=0.48\linewidth]{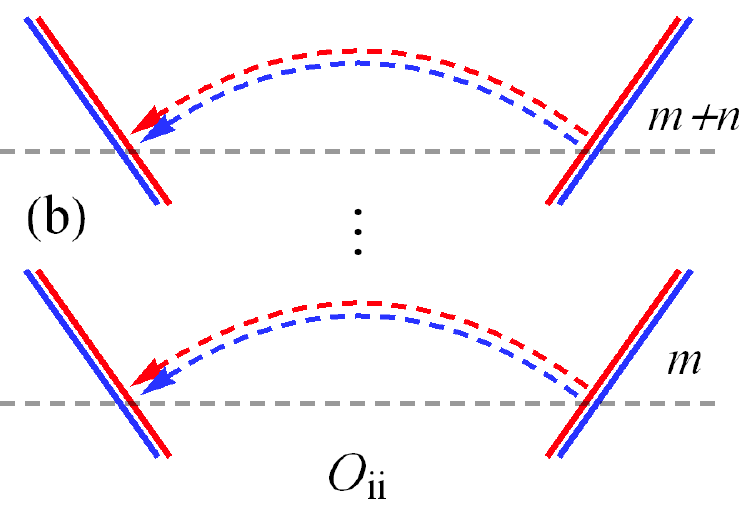} \\
 \includegraphics[width=0.48\linewidth]{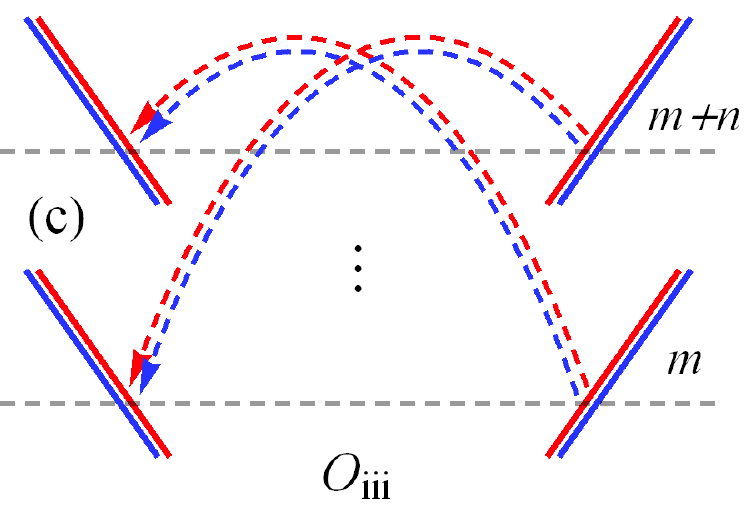}
 \hspace{0.01\linewidth}
 \includegraphics[width=0.48\linewidth]{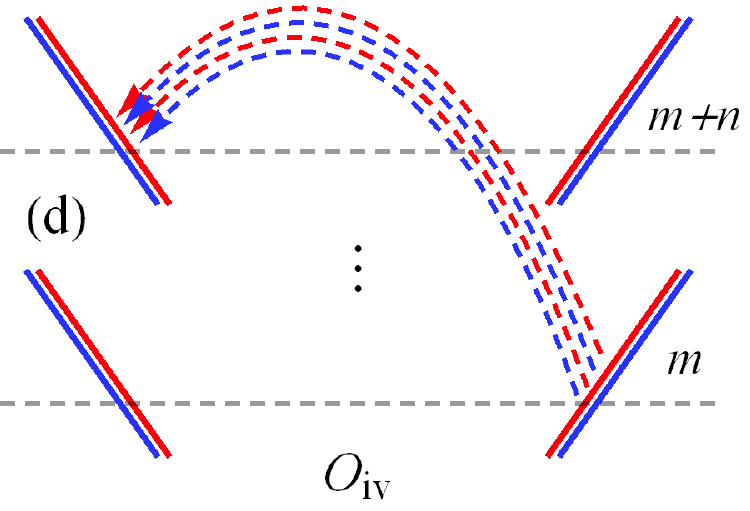}
 \caption{Examples for \moire umklapp scatterings at $\nu = P/4$. 
(a) $O_{\rm i}$, characterized by Eq.~\eqref{Eq:O_gs} with $ (s_{R0\sigma},s_{L0\sigma}) = (2,-2)$.
(b) $O_{\rm ii}$, with $ (s_{R0\sigma},s_{L0\sigma},s_{Rn\sigma},s_{Ln\sigma}) = (1,-1,1,-1)$.
(c) $O_{\rm iii}$, with $ (s_{R0\sigma},s_{L0\sigma},s_{Rn\sigma},s_{Ln\sigma}) = (1,-1,1,-1)$.
(d) $O_{\rm iv}$, with $ (s_{R0\sigma},s_{L0\sigma},s_{Rn\sigma},s_{Ln\sigma}) = (2,0,0,-2)$.
Here we illustrate processes that are invariant upon changing the spin sign; see 
 Table~\ref{Tab:umklapp} 
for more general cases.
}
 \label{Fig:type-II}
\end{figure}

{\it \Moire correlated states.} The \moire umklapp scatterings can be further categorized into four types, depending on whether they involve multiple wires, whether they involve scatterings between wires, and whether they conserve the particle number for each wire. 
While the operator in Eq.~\eqref{Eq:O_gs} describes general processes at fractional fillings in Eq.~\eqref{Eq:nu_condition}, below we provide specific examples allowed at $\nu = P/4$. 

We start with processes involving only single wires and denote the corresponding operator as $O_{\rm i} $. 
In Fig.~\ref{Fig:type-II}(a), we illustrate the process with nonzero coefficients, 
$ (s_{R0\sigma},s_{L0\sigma}) = (2,-2)$ for both $\sigma = \; \uparrow$ and $\sigma = \; \downarrow$.
The example describes a process where four electrons at $k_F$ are backscattered to $-k_F$, with the total momentum difference $8k_F = 4 \nu \times (2\pi / \lambda)$ compensated by the \moire potential.  
Next, there are umklapp processes involving multiple wires with correlated intrawire scatterings, labeled as $O_{\rm ii} $.   
The simplest case involves two $n$-th nearest neighboring wires, with an example in 
Fig.~\ref{Fig:type-II}(b); we note that the number of backscatterings in each wire can be different. 
Furthermore, we have $O_{\rm iii}$ involving interwire scatterings while still conserving the particle number for each wire. 
As mentioned above, the latter constraint implies $\bar{S}_{p,c} = 0 $ for any $p$, as in the case for $O_{\rm i} $ and $O_{\rm ii} $. 
For instance, in Fig.~\ref{Fig:type-II}(c) we show a process involving two $n$-th nearest neighbor wires. 
Finally, allowing for processes which do not conserve the particle number for some wires, we have  $O_{\rm iv}$, with $\bar{S}_{p,c} \neq 0$ for some $p$. 
In Fig.~\ref{Fig:type-II}(d) we plot a two-wire process. 
In addition to the depicted examples, we present the \moire umklapp scatterings in 
Table~\ref{Tab:umklapp} 
in SM~\cite{SM}, covering a broader range of fillings and higher-order processes.

For $O_{\rm i}$, $O_{\rm ii}$, and $O_{\rm iii}$, one can obtain a sum of sine-Gordon terms upon bosonization. Taking Fig.~\ref{Fig:type-II}(a) as an example, we have
$ O_{\rm i} + O_{\rm i}^{\dagger}  \propto \sum_{m} \cos \big( 4 \sqrt{2} \phi_{cm}  \big)$. 
When the corresponding operator is RG relevant, it gaps out all the $\phi_{cm}$ fields and leads to a correlated insulating state at fractional fillings. 
In the strong-coupling limit, $\phi_{cm}$ is pinned to a minimum of the cosine. 
A kink excitation corresponds to a tunneling process between two neighboring minima, where $\phi_{cm}$ changes its value by $\pm \pi/(2\sqrt{2} )$. 
We find that the system hosts fractional excitations with charge $\pm e/ 2$ associated with the kink. 
In contrast to the first three types, the states resulting from $O_{\rm iv}$ can host gapless edge modes, which we demonstrate next.

{\it Chiral edge modes.} 
We consider $O_{\rm iv}$ scattering involving the $n$-th nearest neighbor wires, which allows us to keep only a few nonzero coefficients in Eq.~\eqref{Eq:S_coefficient}; i.e., $S_{n,c} = S_{0,c}$ and $\bar{S}_{n,c}=-\bar{S}_{0,c}$. 
To proceed, we introduce chiral fields $\Phi_{\ell m} = - \ell \phi_{cm} + f \theta_{cm}$
with $f  = - \bar{S}_{0,c} / S_{0,c} $, which satisfy 
\begin{align}
 \big[\Phi_{\ell m} (x),  \Phi_{\ell' m'} (x') \big] =& i \ell \pi  \delta_{\ell \ell'} \delta_{mm'} f \, {\rm sign} (x-x'). 
\label{Eq:commutator}
\end{align} 
The transformation leads to
\begin{align}
 O_{\rm iv} + O_{\rm iv}^{\dagger}  \propto & \sum_{m=1}  \cos \Big\{ \frac{ S_{0,c} }{ \sqrt{2} } \big[ \Phi_{L (m+n)} - \Phi_{R m} \big] \Big\}. 
\label{Eq:Oiv-perturbation} 
\end{align}
The expression indicates the presence of $n$ gapless chiral modes $\Phi_{L,1}, \cdots, \Phi_{L,n}$ at one edge and, similarly, $n$ gapless right-moving modes at the opposite edge. 
To proceed, we define  $\tilde{\Phi}_{m,n} = [\Phi_{L (m+n)} - \Phi_{R m}] /2$ and get 
$O_{\rm iv} + O_{\rm iv}^{\dagger}  \propto \sum_{m=1}  \cos \big(  \sqrt{2}  S_{0,c}  \tilde{\Phi}_{m,n} \big)$.
Using Eq.~\eqref{Eq:commutator}, it can be shown that the $\tilde{\Phi}_{m,n}$ fields for any $m$ commute~\cite{SM}, gapping out the bulk modes in the interior of the system.
Similar to the correlated states induced by $O_{\rm i}$--$O_{\rm iii}$, the system hosts fractional excitations, with charge $ \pm 2e/ S_{0,c}$. 
We expect formation of chirality domains, hosting gapless chiral modes at domain walls~\cite{SM}.
While the formation of domain walls costs energy, (disorder-induced) local magnetic moments can trigger their formation, which increases the entropy and therefore lowers the free energy at finite temperatures.
Remarkably, a finite magnetic field is required to train domains in order to stabilize edge modes with a definite chirality in micrometer-size samples~\cite{Sharpe:2019,Serlin:2020}.

Using the Landauer-B{\"u}ttiker formalism~\cite{Landauer:1957,Landauer:1970,Buttiker:1988,Datta:1995}, we obtain quantized Hall resistance $h/(ne^2)$. 
For $n=1$ and $\nu = 3/4$, it leads to a value of $h/e^2$, as observed in Ref.~\cite{Serlin:2020}. 
In consequence, the system exhibits quantum anomalous Hall effect with chiral edge modes and fractional excitations. 
We note that it is possible to reproduce a sequence of Chern insulating states with $C = \pm 1, \pm 2$ and $\pm 3$ (corresponding to $n$ here) at fillings $\nu = \pm 3/4, \pm1/2$ and $\pm 1/4$, respectively. The complete sequence was observed in Refs.~\cite{Nuckolls:2020,Choi:2021,Das:2021}, while a partial set was reported in Refs.~\cite{Sharpe:2019,Serlin:2020,Saito:2021b,Stepanov:2021,Park:2021b,Lin:2022,Tseng:2022}.

To demonstrate that $O_{\rm iv}$ can be RG relevant, we compute its scaling dimension and get~\cite{SM}
\begin{eqnarray}
\Delta_{\rm iv} 
&=& \frac{1}{2} \left| S_{0,c} \bar{S}_{0,c} \right| \Big(1 + \frac{2 U }{ \hbar v_0} \Big)^{-\frac{1}{2}}, 
\label{Eq:scaling-dim}
\end{eqnarray}
with the $q \sim 0$ Fourier component $U$ of the density-density interaction and the velocity $v_0$. 
In consequence, for a given scattering process, the RG relevance condition $\Delta_{\rm iv} <2$ is fulfilled for sufficiently large $U$.

\begin{figure}[t]
\centering
 \includegraphics[width=0.48\linewidth]{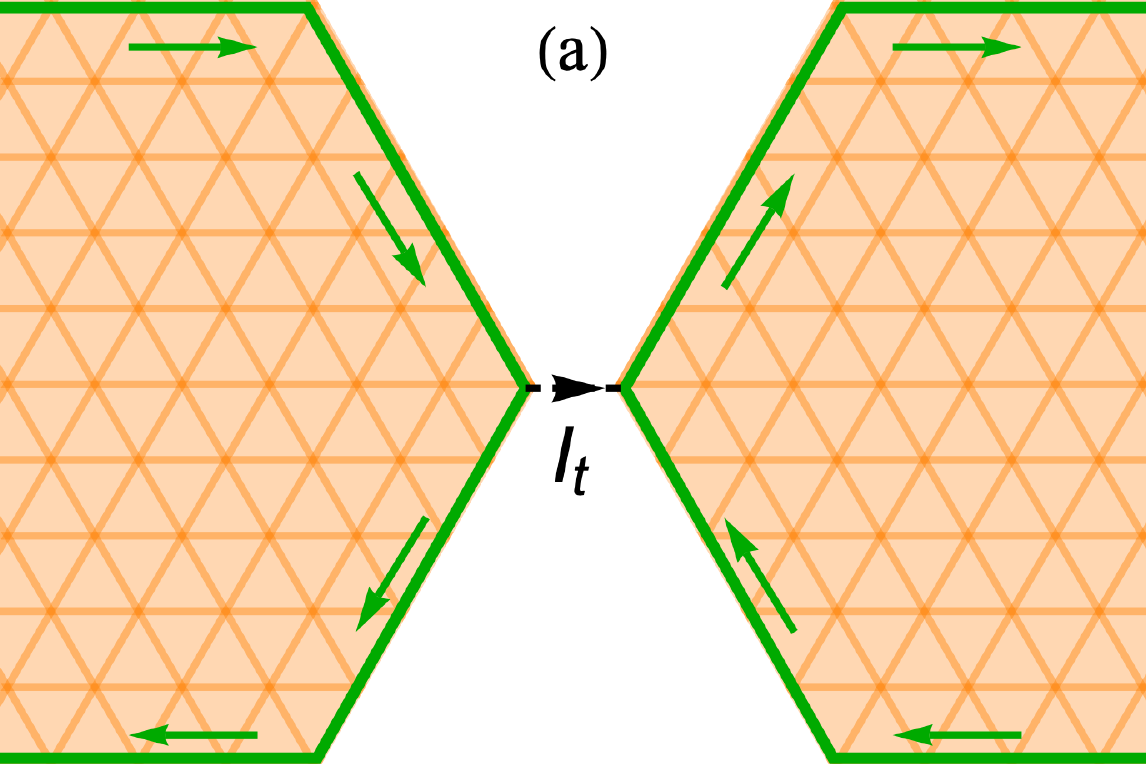}
 \hspace{0.01\linewidth}
 \includegraphics[width=0.48\linewidth]{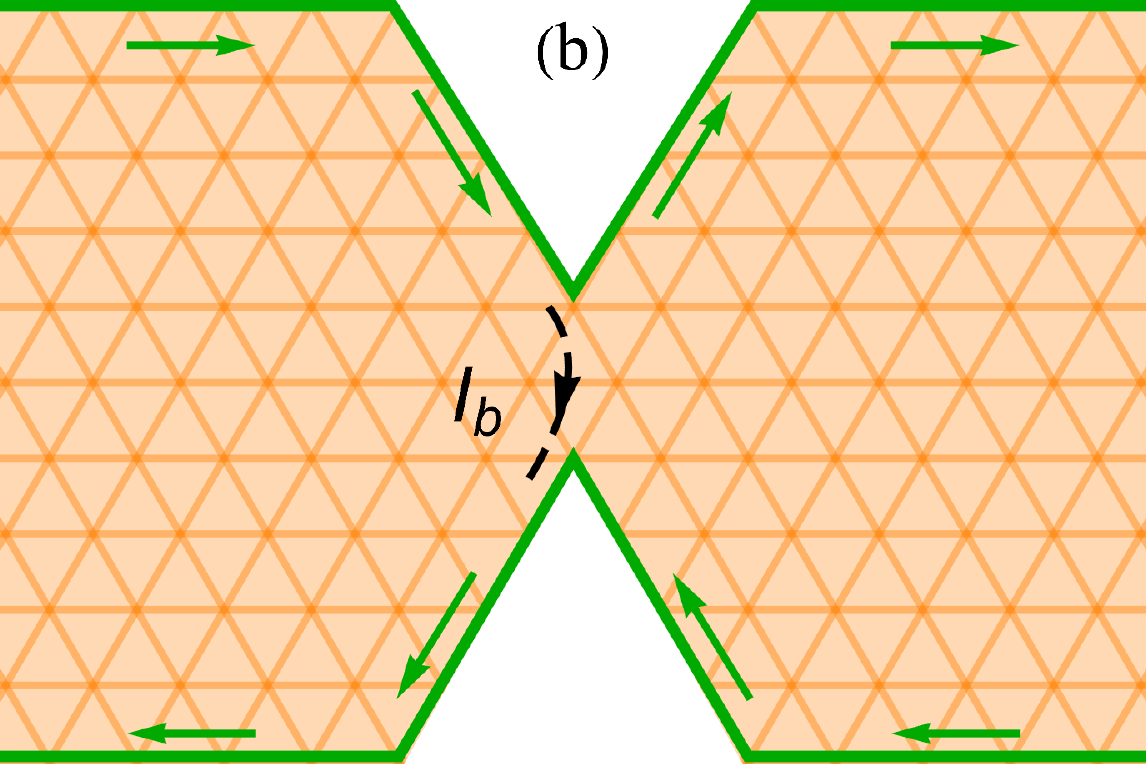}
 \caption{QPC setups for systems in a \moire correlated state with a gapped bulk (orange) and gapless chiral edge modes (green). The setup (a) allows for tunnel current $I_{\rm t}$. The setup (b) leads to backscattering current $I_{\rm b}$ and conductance correction $\delta G$.
}
 \label{Fig:QPC}
\end{figure}

{\it Experimental signatures.} The predicted chiral edge modes can be characterized by spectroscopic and transport measurements. 
For simplicity we consider a \moire correlated state hosting a single edge mode~\cite{SM}.
Utilizing scanning tunneling spectroscopy, one can probe the local density of states, which follows a universal scaling curve for energy $\epsilon$ and temperature $T$,
\begin{eqnarray}
\rho (\epsilon) &\propto& T^{ \frac{1}{f} -1} \cosh \left( \frac{ \epsilon}{2k_{\rm B} T} \right) \left| \Gamma \left( \frac{ 1}{2f} + i \frac{ \epsilon }{2 \pi k_{\rm B} T} \right)\right|^2 . \hspace{12pt} 
\label{Eq:DOS} 
\end{eqnarray}
In contrast to carbon nanotubes~\cite{Bockrath:1999,Balents:1999} or helical liquids~\cite{Stuhler:2019,Hsu:2021}, the scaling exponent here does not depend on  $H_{\rm fs}$, demonstrating the topological nature of the chiral edge modes.  

Alternatively, one can probe the chiral edge modes via charge transport~\cite{Kane:1994,Kane:1995,Kane:1996,Fisher:1997,Chang:2003,Hsu:2019}. Specifically, we consider two setups employing quantum point contacts (QPCs). 
The setup in Fig.~\ref{Fig:QPC}(a) allows for interedge tunneling with a current described by another universal scaling formula, 
\begin{eqnarray}
I_{\rm t} &\propto& T^{  \frac{2}{f} -1} \sinh\left( \frac{ eV}{2k_{\rm B} T} \right) \left| \Gamma \left( \frac{ 1}{f} + i \frac{eV}{2 \pi k_{\rm B} T} \right)\right|^2,  
\label{Eq:I-V} 
\end{eqnarray}
with bias voltage $V$. 
On the other hand, the interedge backscattering in Fig.~\ref{Fig:QPC}(b) leads to power-law correction in the (differential) conductance with the magnitude $| \delta G| \propto {\rm max} (eV, k_{\rm B}T)^{2f -2}$. Unlike fractional quantum Hall states~\cite{Kane:1994,Kane:1995,Kane:1996,Fisher:1997,Chang:2003}, the scaling exponents depend on $f$ here, but not directly on the filling factor $\nu$. 
 The same QPC geometry can be used to detect fractional charges through shot noise~\cite{Kane:1994a,Saminadayar:1997,dePicciotto:1997}. 

As a remark, while the theoretical works establishing the quantum-wire network in \moire bilayers~\cite{San-Jose:2013,Efimkin:2018} involve a sufficiently large interlayer potential difference, achievable via voltage gates, we expect that the network can form under broader conditions~\cite{Huang:2018,Rickhaus:2018,Choi:2019,Kerelsky:2019,Jiang:2019,Xie:2019}. 
Namely, a spectral gap can be generally induced in graphene-based devices through coupling to other layered materials or substrates, depending on their stacking configurations~\cite{Dean:2013,Hunt:2013,Ponomarenko:2013,Moon:2014,Kim:2018,Sharpe:2019,Serlin:2020,Lin:2022,Tseng:2022}. Therefore, a gap with a spatially dependent sign can be achieved through nanoscale engineering~\cite{Kindermann:2012,Hsu:2020n,Chen:2020}, leading to a network of gapless domain walls that separate regions with opposing gap signs.

Finally, we point out that, through the proposed experimental verification, the system can reveal the long-sought intrinsic fractional quantum anomalous Hall states, where topology and many-body physics interplay. 
Upon inducing superconductivity (e.g., by proximity), \moire correlated states hosting fractional edge modes provide a platform to stabilize parafermion edge or zero modes~\cite{Fendley:2012,Clarke:2013,Klinovaja:2014c,Mong:2014,Meng:2014b,Oreg:2014,Sagi:2015,Sagi:2017,Laubscher:2019,Laubscher:2020} even without magnetic fields.


\begin{acknowledgments}
{\it Acknowledgments.} 
We thank Y.-Y. Chang, C.-H. Chung, and C.-T. Ke for interesting discussions. 
This work was financially supported by the JSPS Kakenhi Grant No.~19H05610, the Swiss National Science Foundation (Switzerland), the NCCR QSIT, and the National Science and Technology Council (NSTC), Taiwan through NSTC-112-2112-M-001-025-MY3.
\end{acknowledgments}


\bibliography{arxiv}

\clearpage
\onecolumngrid

\bigskip

\begin{center}
\large{\bf Supplemental Material to ``General scatterings and electronic states in the quantum-wire network of \moire systems'' }

\vspace{6pt} 
\fontsize{10}{12}
Chen-Hsuan Hsu$^{1,2,3}$, Daniel Loss$^{3,4}$, and Jelena Klinovaja$^{4}$ \\
\vspace{4pt} 
{\it
$^{1}$Yukawa Institute for Theoretical Physics, Kyoto University, Kyoto 606-8502, Japan \\
$^{2}$Institute of Physics, Academia Sinica, Taipei 115, Taiwan \\
$^{3}$RIKEN Center for Emergent Matter Science, Wako, Saitama 351-0198, Japan \\
$^{4}$Department of Physics, University of Basel, Klingelbergstrasse 82, CH-4056 Basel, Switzerland 
}
\end{center}

\twocolumngrid
\setcounter{equation}{0}
\setcounter{figure}{0}
\setcounter{table}{0}
\setcounter{page}{1}
\setcounter{NAT@ctr}{0}   
\makeatletter
\renewcommand{\theequation}{S\arabic{equation}}
\renewcommand{\thefigure}{S\arabic{figure}}
\renewcommand{\thetable}{S\arabic{table}}
\renewcommand{\bibnumfmt}[1]{[S#1]}
\renewcommand{\citenumfont}[1]{S#1}

\subsection{I. Conventional scatterings}
In this section we discuss conventional scatterings, which fulfill Eq.~\eqref{Eq:conditionLR2} in the main text and can take place at any fillings. This category includes charge-density-wave (CDW) couplings, Josephson couplings, and hoppings, corresponding to the ``crystalline states'' discussed in spinless fermion systems~\citesupp{Kane:2002_S}.

\begin{figure}[h]
\centering
 \includegraphics[width=0.48\linewidth]{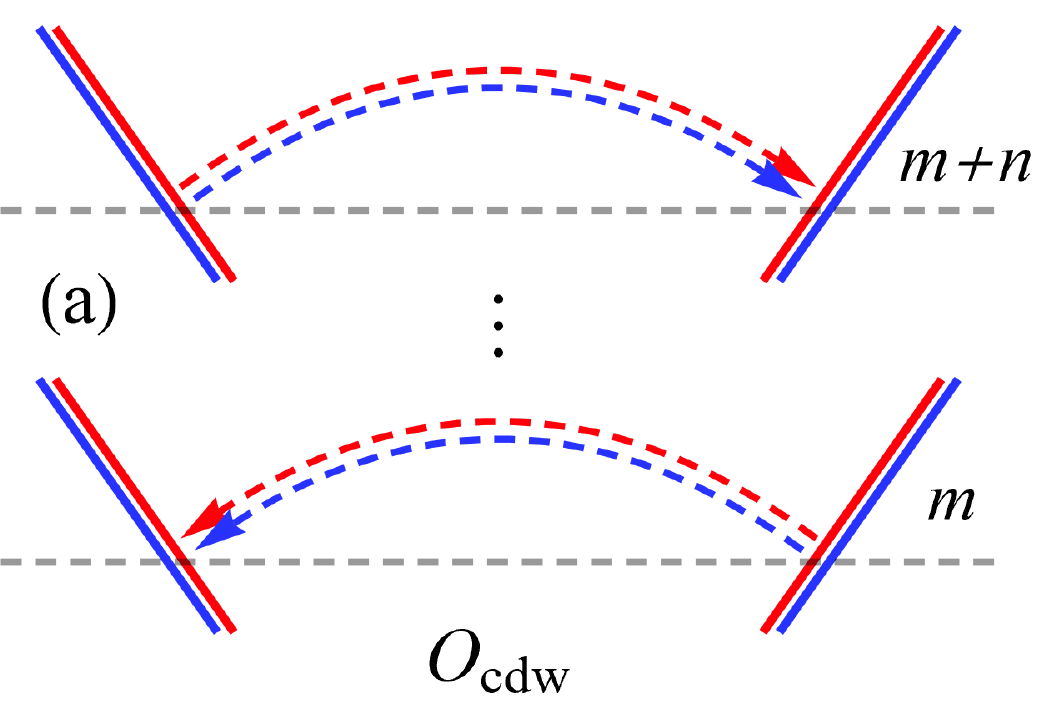}
 \hspace{0.01\linewidth}
 \includegraphics[width=0.48\linewidth]{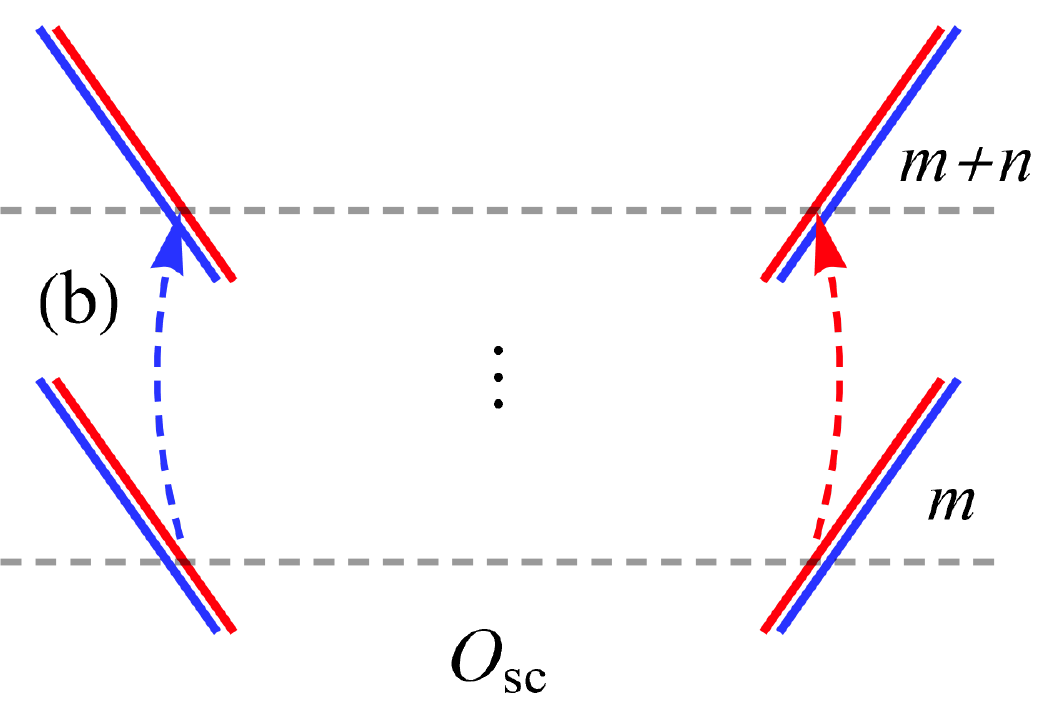}\\
 \includegraphics[width=0.48\linewidth]{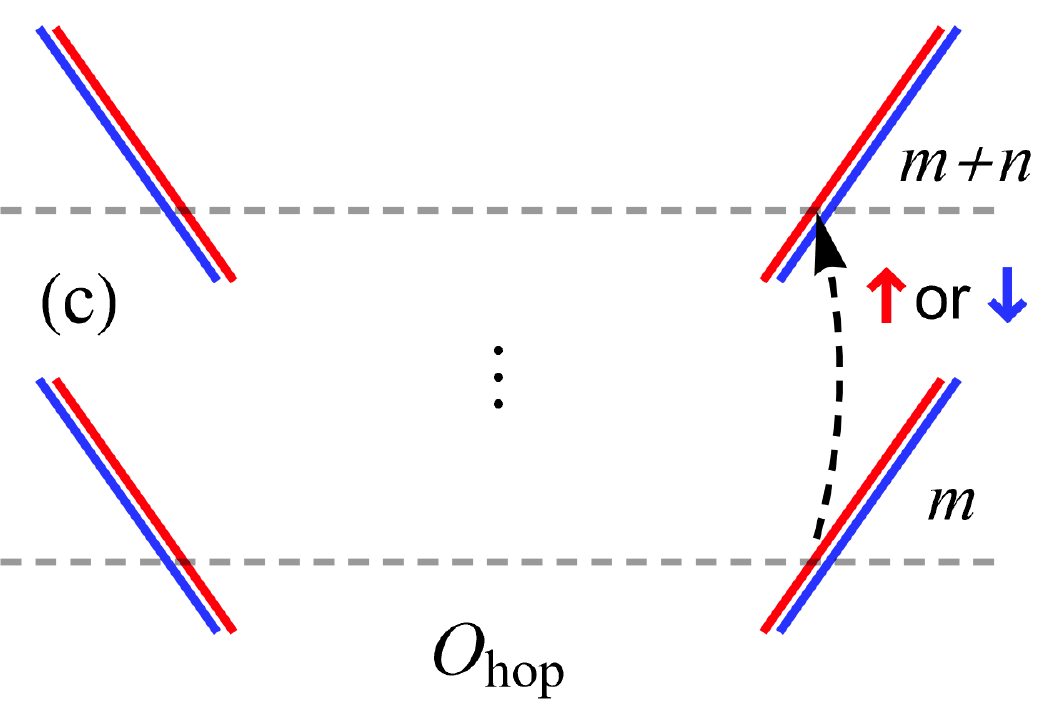}
 \caption{Illustrations of conventional scattering processes, which can take place at any fillings. 
(a) CDW coupling, characterized by Eq.~\eqref{Eq:O_cdw} with $(N_{0\uparrow},N_{0\downarrow},N_{n\uparrow},N_{n\downarrow})=(-1,-1,1,1)$.
(b) Josephson coupling, characterized by Eq.~\eqref{Eq:O_sc} with $(N_0,M_0,N_n,M_n) = (0,-1,0,1)$.
(c) Single-particle hopping, characterized by Eq.~\eqref{Eq:O_gs} with $(s_{R0 \sigma}, s_{R n \sigma}, s_{L0 \sigma}, s_{L n \sigma} ) = (-1,1,0,0)$ for $\sigma = \uparrow$ or $\downarrow$. 
}
 \label{Fig:type-I}
\end{figure}

We start with the generalized CDW couplings, which include higher-order processes involving a single wire or multiple wires.
For simplicity, we focus on processes with $\bar{S}_{p,c}=0$ for any $p$. 
It allows us to rewrite the coefficients $s_{\ell p \sigma}$ with integers $N_{p\sigma}$, 
\begin{eqnarray}
\hspace{-6pt} 
(s_{Rp\uparrow},s_{Lp\uparrow},s_{Rp\downarrow},s_{Lp\downarrow}) = (-N_{p\uparrow},N_{p\uparrow},-N_{p\downarrow},N_{p\downarrow}),
\label{Eq:O_cdw}
\end{eqnarray}
which fulfill the global particle number conservation.
Furthermore, the integer set fulfills the momentum conservation when $\sum_{p} (N_{p\uparrow}+N_{p\downarrow}) = 0 $. 
The operator can be bosonized in the form of Eq.~\eqref{Eq:O_general} in the main text with $ S_{p,\xi} = 2( N_{p\uparrow} +\xi N_{p\downarrow})$ and $ \bar{S}_{p,\xi} = 0$.
As an example, in Fig.~\ref{Fig:type-I}(a) we illustrate the two-wire scattering process with $(N_{0\uparrow},N_{0\downarrow},N_{n\uparrow},N_{n\downarrow})=(-1,-1,1,1)$ and $N_{p\sigma} = 0$ for $p \neq 0,n$.
It is straightforward to check that, for sufficiently strong electron-electron interaction, the CDW operator is RG relevant, leading to the CDW phase with an insulating bulk.

In addition, the generalized Josephson couplings (assuming singlet pairing) can be characterized by  integers $M_{p}$ and $N_{p}$,
\begin{eqnarray}
\label{Eq:O_sc}
(s_{Rp\uparrow},s_{Lp\uparrow},s_{Rp\downarrow},s_{Lp\downarrow}) = (-M_{p},N_{p},N_{p},-M_{p}),
\end{eqnarray}
which already incorporates the momentum conservation condition. Moreover, the global particle number conservation is ensured by $\sum_{p} (N_{p} - M_{p}) = 0 $. 
The operator can be bosonized in the form of Eq.~\eqref{Eq:O_general} in the main text with  
$\bar{S}_{p,c} = 2 (N_p - M_p)$, $S_{p,s} = 2 (N_p + M_p)$, $S_{p,c} = 0$, and $\bar{S}_{p,s}=0$.
When the corresponding operator is RG relevant, the system is in the superconducting state. 
An example with $(N_0,M_0,N_n,M_n) = (0,-1,0,1)$  and $N_{p}, M_{p} = 0$ for $p \neq 0,n$. is illustrated in Fig.~\ref{Fig:type-I}(b).

There are also generalized hopping processes, including single-particle hopping and pair hopping between wires. In contrast to the above, the expression for this type of scatterings cannot be simplified; we therefore keep the general notation $s_{\ell p \sigma}$ in Eq.~\eqref{Eq:O_gs}.
If the operator is RG relevant, it can lead to a two-dimensional Fermi liquid~\citesupp{Emery:2000_S,Vishwanath:2001_S,Mukhopadhyay:2001a_S,Mukhopadhyay:2001b_S}. 
An example with $(s_{R0 \sigma}, s_{R n \sigma}, s_{L0 \sigma}, s_{L n \sigma} ) = (-1,1,0,0)$  is illustrated in Fig.~\ref{Fig:type-I}(c).

\begin{table*}[t]
\caption{\Moire umklapp scatterings. The operators, $O_{\rm i}$--$O_{\rm iv}$, are in the fermion form in Eq.~\eqref{Eq:O_gs} with the listed $ s_{\ell p\sigma} $ values and 
$\mathbb{N}$ denoting the positive integer set. 
The scatterings are allowed at the listed filling factor $\nu$, with $P$ being a nonzero integer.
The bosonized form is in Eq.~\eqref{Eq:O_general} with the listed $S_{p,\xi}$ and $\bar{S}_{p,\xi} $ values. 
While we list two-wire processes for $O_{\rm ii}$--$O_{\rm iv}$ for simplicity, processes involving more wires are generally allowed as long as Eqs.~\eqref{Eq:ChargeConser}--\eqref{Eq:MomentumConser} are fulfilled. }
\begin{center}
\begin{tabular}{c c c c c c }
\hline \\
[-8pt]  
Operator & $ s_{\ell p\sigma} $ & possible values & $\nu$ & $S_{p,\xi}$ &  \vspace{2pt}  $    \bar{S}_{p,\xi} $ \\
\hline
$O_{\rm i}$ & $\ell \delta_{p0} N_{\sigma}$ &
$N_{ \sigma} \in \mathbb{N}$ & 
$ \frac{P}{  \sum_{\sigma} N_{\sigma}  }$   
& 
$ -2 \delta_{p0} (N_{\uparrow} + \xi N_{\downarrow}) $ & 
0 \\ [6pt]
$O_{\rm ii}$ & $\ell (\delta_{p0} + \delta_{pn}) N_{p \sigma}  $ &
$N_{0 \sigma}$, $N_{n \sigma} \in \mathbb{N} $  & 
$ \frac{P}{  \sum_{\sigma}  (N_{0\sigma}  + N_{n\sigma}  )  }$  
& 
$  -2 (\delta_{p0} + \delta_{pn} ) (N_{p\uparrow} + \xi N_{p\downarrow}) $ & 
0 \\ [6pt]
$O_{\rm iii}$ & $\ell (\delta_{p0} + \delta_{pn}) N_{ \sigma}  $ &
$N_{ \sigma} \in \mathbb{N}$ & 
$ \frac{P}{  2 \sum_{\sigma} N_{\sigma}  }$  
& 
$ -2 (\delta_{p0} + \delta_{pn} ) (N_{\uparrow} + \xi N_{\downarrow}) $ & 
0 \\ [6pt]
$O_{\rm iv}$ & 
\begin{tabular}{l}
$\delta_{\ell R} (\delta_{p0} N_{ \sigma}  + \delta_{pn} M_{ \sigma} )$ \\
$  - \delta_{\ell L} (\delta_{p0} M_{ \sigma}  + \delta_{pn} N_{ \sigma} )$
\end{tabular} &
 $N_{ \sigma}$, $M_{ \sigma} \in \mathbb{N} $, $N_{ \sigma} \neq M_{ \sigma}$
&
$ \frac{P}{  \sum_{\sigma}  ( N_{\sigma} + M_{\sigma} ) }$  
& 
\begin{tabular}{l}
$ -2 (\delta_{p0} + \delta_{pn} ) \delta_{\xi c} (N_{\uparrow} + M_{\downarrow}) $ \\
$ -2 (\delta_{p0} + \delta_{pn} ) \delta_{\xi s} (N_{\uparrow} - N_{\downarrow}) $
\end{tabular}
& 
$ 2 (\delta_{p0} - \delta_{pn} ) \delta_{\xi c} (N_{\uparrow} - M_{\uparrow}) $ \\
\hline 
\end{tabular}
\end{center}
\label{Tab:umklapp}
\end{table*}


\subsection{II. \Moire umklapp scatterings  }
In this section, we summarize the four types of \moire umklapp scatterings, before discussing the scattering described by $O_{\rm iv}$ in more detail. 
Aiming at a systematic construction of the scattering operators, for each of the types we will introduce positive integer coefficients, in order to decrease the number of independent coefficients. 
We start with processes involving only single wires and denote the corresponding operator as $O_{\rm i} $. In the form of Eq.~\eqref{Eq:O_gs}, $O_{\rm i} $ is characterized by the following nonzero $s_{\ell p\sigma}$,
\begin{eqnarray}
\hspace{-6pt} 
(s_{R0\sigma},s_{L0\sigma}) = (N_{\sigma},-N_{\sigma}),
\label{Eq:O_um1}
\end{eqnarray}
with positive integers $N_{\sigma}$.
 
Next, there are umklapp processes involving multiple wires with correlated intrawire scatterings.   
The simplest case involves two $n$-th nearest neighboring wires, labeled as $O_{\rm ii} $, with nonzero $s_{\ell p\sigma}$,
\begin{eqnarray}
\hspace{-12pt} 
(s_{R0\sigma},s_{L0\sigma},s_{Rn\sigma},s_{Ln\sigma}) = (N_{0\sigma},-N_{0\sigma},N_{n\sigma},-N_{n\sigma}),
\label{Eq:O_um2}
\end{eqnarray}
with positive integers $N_{0\sigma}$ and $N_{n\sigma}$.
 
Furthermore, we consider \moire umklapp scatterings with interwire processes while still conserving the particle number for each wire. The latter condition implies $\bar{S}_{p,c} = 0 $ for any $p$. 
Together with the choice $\bar{S}_{p,s} = 0 $ and limiting ourselves to processes involving two $n$-th nearest neighboring wires, we get $O_{\rm iii}$ with nonzero $s_{\ell p\sigma}$ given by
\begin{eqnarray}
(s_{R0\sigma},s_{L0\sigma},s_{Rn\sigma},s_{Ln\sigma}) = (N_{\sigma},-N_{\sigma},N_{\sigma},-N_{\sigma}),
\label{Eq:O_um3}
\end{eqnarray}
with positive integers $N_{\sigma}$.

Finally, allowing for processes which do not conserve the particle number for some wires, we have $\bar{S}_{p,c} \neq 0$ for some $p$. 
Taking two-wire processes for simplicity and $\bar{S}_{p,s} = 0 $, we have $O_{\rm iv}$ as Eq.~\eqref{Eq:O_gs} with 
\begin{eqnarray}
(s_{R0\sigma},s_{L0\sigma},s_{Rn\sigma},s_{Ln\sigma}) = (N_{\sigma},-M_{\sigma},M_{\sigma},-N_{\sigma}),
\label{Eq:O_um4}
\end{eqnarray}
with positive integers $N_{\sigma}$ and $M_{\sigma}$.
Here $N_{\sigma} \neq M_{\sigma}$, as otherwise we would have processes already included in $O_{\rm iii}$. 
By introducing indices such as $N_{\sigma}$ and $M_{\sigma}$, we can decrease the number of independent indices within a scattering subtype, enabling a systematic construction of the scattering operators. We summarize the \moire umklapp scatterings in Table~\ref{Tab:umklapp}, which includes higher-order processes for general filling factors.

Owing to their possibility for hosting chiral edge modes, we analyze the \moire umklapp scattering described by $O_{\rm iv}$ in more detail.  
As in the main text, we have   
\begin{eqnarray}
H_{\rm iv} = \sum_{m} g_{\rm iv} \int dx \; \cos \big( \sqrt{2} S_{0,c}  \tilde{\Phi}_{m,n} \big),
\label{Eq:Hiv}
\end{eqnarray}
with the coupling parameter $g_{\rm iv}$ and the transformed field $\tilde{\Phi}_{m,n}$ introduced in the main text. Here, we have further simplify our analysis by considering processes with $N_{\uparrow} = N_{\downarrow}$, i.e., $S_{p,s}=0$.
With the coefficients in Eq.~\eqref{Eq:O_um4}, we have $S_{0,c} = -( N_{\uparrow} + N_{\downarrow} + M_{\uparrow} + M_{\downarrow} )$ and $f  = (N_{\uparrow} - M_{\uparrow}) / (N_{\uparrow} + M_{\downarrow}) $. 
Using the commutator in Eq.~\eqref{Eq:commutator}, it can be shown that the transformed field fulfills the relation,
\begin{eqnarray}
\big[\tilde{\Phi}_{m,n}(x),  \tilde{\Phi}_{m',n} (x') \big] &=& 
\frac{1}{4} \Big( \big[ \Phi_{L,m+n}(x),  \Phi_{L,m'+n} (x') \big] \nonumber \\
&& \hspace{5pt} + \big[ \Phi_{R,m}(x),  \Phi_{R,m'}(x') \big] 
\Big) \nonumber \\
&=& ( -1 + 1 ) \frac{ \pi i f }{4} \delta_{mm'} {\rm sign} (x-x')  \nonumber \\
&=& 0 .
\end{eqnarray}
Since the $\tilde{\Phi}_{m,n}$ fields for all $m$ commute with each other, the cosine terms in Eq.~\eqref{Eq:Hiv} can be ordered simultaneously, gapping out the bulk modes.
Introducing $\tilde{\Theta}_{m,n} = (\Phi_{L (m+n)} + \Phi_{R m})/2$, we obtain the additional relations,
\begin{eqnarray}
 &&  \big[\tilde{\Theta}_{m,n}(x),  \tilde{\Theta}_{m',n} (x') \big] = 0 , \nonumber \\
 && \big[\tilde{\Phi}_{m,n}(x),  \tilde{\Theta}_{m',n} (x') \big] =\frac{i \pi }{2} f \delta_{mm'} {\rm sign} (x'-x).
\end{eqnarray}

To proceed, we follow Kane~et~al.~\citesupp{Kane:2002_S} and express the forward-scattering term of the density-density interaction in the transformed basis, where the unperturbed part of the effective action is given by 
\begin{eqnarray}
\frac{ \tilde{S}_{\rm 0}}{\hbar} &=& \sum_m \int \frac{ dx d\tau }{ \pi } \, \left[
\frac{-i}{ f}   (\partial_x \tilde{\Theta}_{m,n}) (\partial_\tau \tilde{\Phi}_{m,n}) \right.  \nonumber \\
&&   \left.  + \Big( \frac{v_0}{2} + \frac{U}{ \hbar}  \Big) \left( \partial_{x} \tilde{\Phi}_{m,n} \right)^2  + \frac{v_0}{2} \left( \partial_{x} \tilde{\Theta}_{m,n} \right)^2
\right]. \nonumber \\
\end{eqnarray}
We note that the fraction $f$ comes from the above commutator and will enter the scaling dimensions of various operators containing $\tilde{\Phi}_{m,n}$ or $\tilde{\Theta}_{m,n}$. 
In the momentum and Matsubara frequency domain, we have
\begin{eqnarray}
\frac{ \tilde{S}_{\rm 0} }{\hbar} &=&  \frac{1}{2\pi \beta \hbar \Omega }\sum_m  \sum_{q,\omega_n} 
\Big( \tilde{\Phi}_{m,n}^{*} (q,\omega_n),  \tilde{\Theta}_{m,n}^{*} (q,\omega_n) \Big) \nonumber \\
&& \times 
\left(
\begin{array}{cc}
v_0 q^2 (1 + \frac{2U} {\hbar v_0} ) & \frac{i}{f} q \omega_n   \\
 \frac{i}{f} q \omega_n  & v_0 q^2 
\end{array}
\right)
\left(
\begin{array}{c}
\tilde{\Phi}_{m,n}(q,\omega_n) \\
\tilde{\Theta}_{m,n} (q,\omega_n)
\end{array}
\right). \nonumber \\
\end{eqnarray}
Inverting the matrix above, one can compute 
$\langle \big[ \tilde{\Phi}_{m,n}(x,\tau)- \tilde{\Phi}_{m,n}(0,0) \big]^2 \rangle$ with respect to $\tilde{S}_{0}$ and obtain the scaling dimension of $O_{\rm iv}$, 
\begin{eqnarray}
\Delta_{\rm iv} &=& \frac{1}{2} |f| S_{0,c}^2 \Big(1 + \frac{2 U}{  \hbar v_0} \Big)^{-1/2},
\end{eqnarray}
which gives Eq.~\eqref{Eq:scaling-dim} in the main text. 
Taking the coefficients in Eq.~\eqref{Eq:O_um4}, one obtains
\begin{eqnarray}
\Delta_{\rm iv} 
&=& 2  (N_{\uparrow}+M_{\downarrow}) \left| N_{\uparrow}-M_{\uparrow} \right| \Big(1 + \frac{2 U }{ \hbar v_0} \Big)^{-\frac{1}{2}}.  
\end{eqnarray}
In consequence, for a given set of $(N_{\sigma},M_{\sigma})$, the scaling dimension $\Delta_{\rm iv} $ can be below 2 for sufficiently large $U$.

\begin{figure}[h]
\centering
	\includegraphics[width=0.46\linewidth]{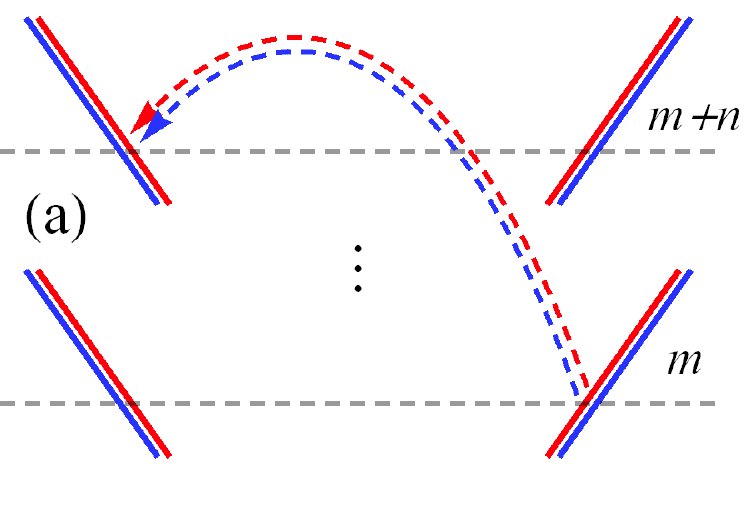}
	\hspace{0.02\linewidth}
 	\includegraphics[width=0.46\linewidth]{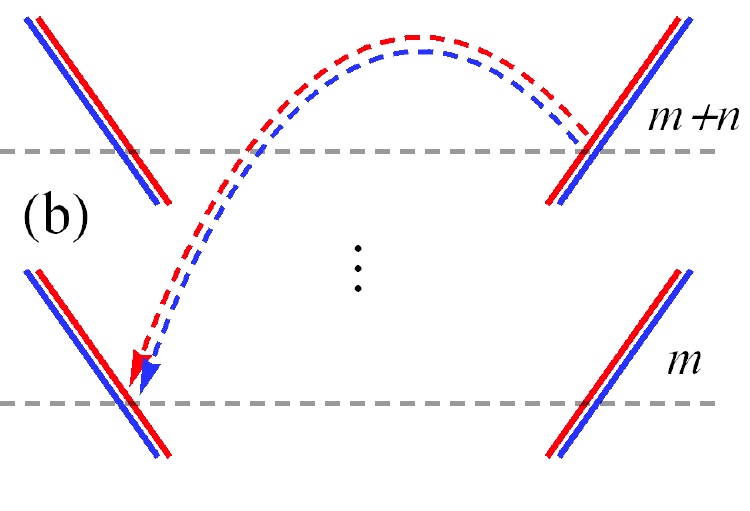}
\caption{$O_{\rm iv}$ processes of the \moire umklapp scatterings at half filling, described by (a) $O_{\rm a}$ in Eq.~\eqref{Eq:O_um_iv-a} and (b) $O_{\rm b}$ in Eq.~\eqref{Eq:O_um_iv-b}.
}
 \label{Fig:um4ab}
\end{figure}

\subsection{III. Chirality domains}
In this section we discuss the $O_{\rm iv}$ processes of the \moire umklapp scatterings and the formation of chirality domains.
For concreteness, we consider the two processes allowed at half filling in Fig.~\ref{Fig:um4ab}; the following discussion can be straightforwardly generalized to other fillings.
The process in Fig.~\ref{Fig:um4ab}(a) is described by
\begin{eqnarray}
\label{Eq:O_um_iv-a}
O_{\rm a} &=& \sum_{m=1} \psi_{L (m+n) \uparrow} ^\dagger \psi_{R m \uparrow} \psi_{L (m+n) \downarrow}^\dagger \psi_{R m \downarrow}  ,   
\end{eqnarray}
corresponding to $O_{\rm iv}  $ with ${N_{\uparrow}} = {N_{\downarrow}} = 1$ and ${M_{\uparrow}} = {M_{\downarrow}} = 0$ [see Eq.~\eqref{Eq:O_um4}]. 
The process in Fig.~\ref{Fig:um4ab}(b) is 
\begin{eqnarray}
\label{Eq:O_um_iv-b}
O_{\rm b} &=& \sum_{m=1} \psi_{L m \uparrow}^\dagger \psi_{R (m+n) \uparrow} \psi_{L m \downarrow} ^\dagger \psi_{R (m+n) \downarrow} , 
\end{eqnarray}
corresponding to ${N_{\uparrow}} = {N_{\downarrow}} = 0$ and ${M_{\uparrow}} = {M_{\downarrow}} = 1$. 
At the half filling, both $O_{\rm a}$ and $O_{\rm b}$ preserve the particle number, spin, and momentum conservation, and therefore fulfill Eqs.~\eqref{Eq:ChargeConser}--\eqref{Eq:MomentumConser} in the main text.

Next, we discuss the stability of the chiral edge modes. Consider the following operators in the bosonized form,
\begin{subequations}
\begin{eqnarray}
O_{\rm a} + O_{\rm a}^\dagger  & \propto & \sum_{m=1}  \cos \Big\{   \sqrt{2}  \big[ \Phi_{L (m+n)} - \Phi_{R m} \big] \Big\}, \hspace{5pt} \\
O_{\rm b} + O_{\rm b}^\dagger & \propto & \sum_{m=1}  \cos \Big\{   \sqrt{2}  \big[ \Phi_{R (m+n)} - \Phi_{L m} \big] \Big\}, \hspace{5pt} 
\end{eqnarray}
\end{subequations}
where we have $ {\rm a} \leftrightarrow  {\rm b}$ upon swapping $R$ and $ L$.
If we are in the parameter regime given in Eq.~\eqref{Eq:scaling-dim} in the main text, where the operators $O_{\rm a, b}$ are RG relevant, their coupling constants can flow to the strong-coupling limit, leading to correlated states with the opposite chirality of the edge modes.

\begin{figure}[t]
\centering
 \includegraphics[width=0.48\linewidth]{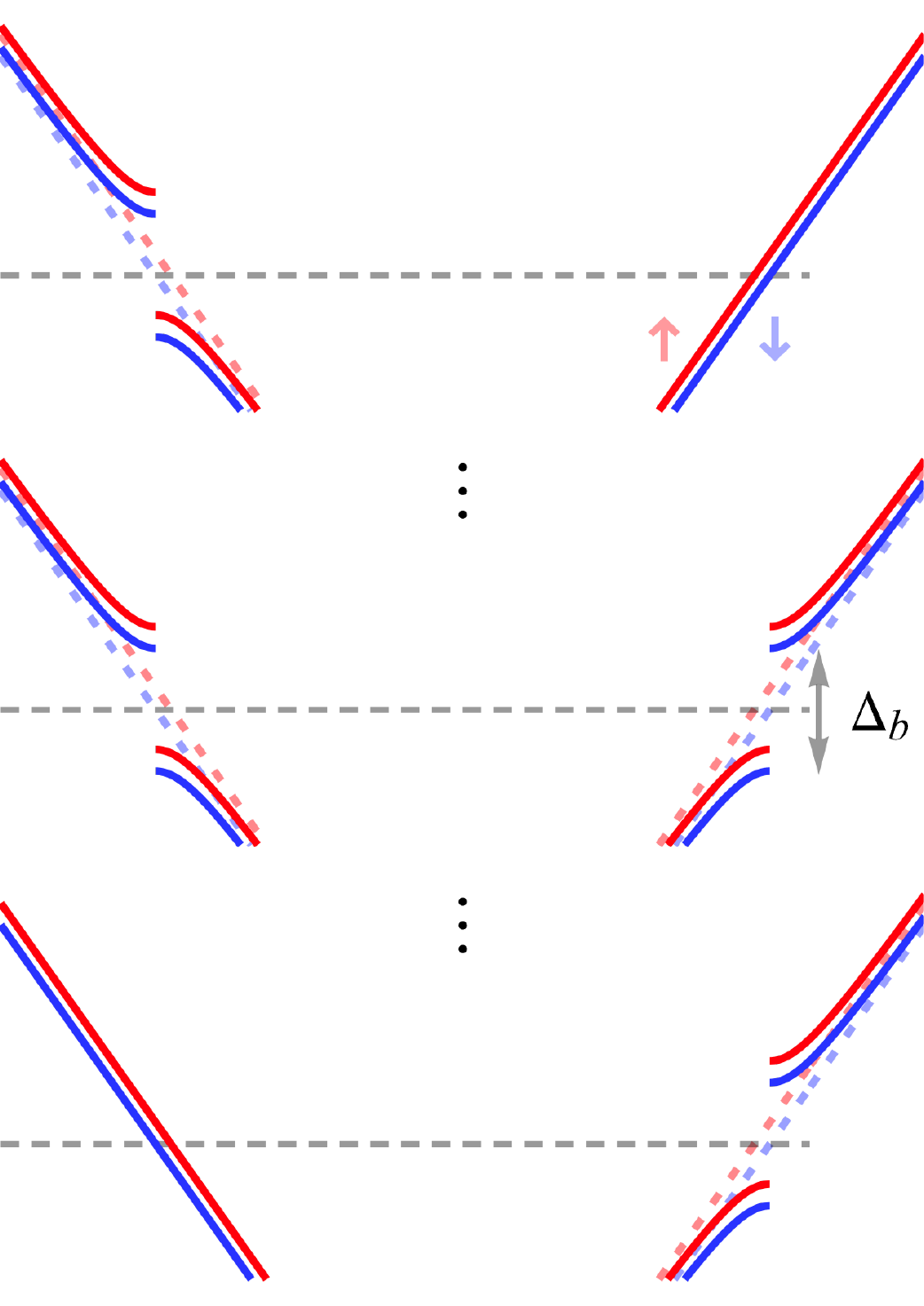}
 \caption{The \moire umklapp scattering $O_{\rm iv}$ induces a bulk gap $\Delta_{b}$ opening in the interior of the system while leaving chiral edge modes gapless.
 The gap opening around the Fermi point $\pm k_F$ leads to a lowering of the energy bands (solid curves) compared to those before the gap opening (dotted curves). This results in an energy gain $\delta \epsilon_{q}$ at momentum $\hbar q$ and the total energy gain per branch given by Eq.~\eqref{Eq:Peierls}.
}
 \label{Fig:Peierls}
\end{figure}

One may wonder if we have the following perturbation,
\begin{eqnarray}
\delta H  &=& \int dx \; \left( g_{\rm a} O_{\rm a} + g_{\rm b} O_{\rm b} + {\rm H.c.} \right),
\label{Eq:Hab}
\end{eqnarray}
whether the edge modes created by a cosine term in $\delta H$ would be gapped out by the other, resulting in a fully gapped system. However, it can be shown that the operators do not commute, 
\begin{eqnarray}
[O_{\rm a} (x) + O_{\rm a}^\dagger (x)  , O_{\rm b} (x') + O_{\rm b}^\dagger (x') ] & \neq& 0,
\end{eqnarray}
so they cannot be ordered simultaneously. 
Therefore, even if $O_{\rm a}$ and $O_{\rm b}$ coexist in $\delta H$ and are both RG relevant, only one of them can open a bulk gap, as shown in Fig.~\ref{Fig:Peierls}.

In each gapped branch, there is an energy gain at momentum $\hbar q$~\citesupp{Meng:2014_S},
\begin{eqnarray}
\delta \epsilon_{q} = \frac {\epsilon_q - \epsilon_{q'}}{2} + \sqrt{ \Big( \frac {\epsilon_q - \epsilon_{q'}}{2}\Big)^2 + \Delta_{b}^2 },
\end{eqnarray}
with the dispersion $\epsilon_q = \hbar v_0 |q|$, $q'= q-2k_F$ for $q\in [0, k_F]$,
$q'=q+2k_F$ for $q\in [-k_F,0]$, and the (renormalized) bulk gap $\Delta_{b}$.
The Peierls energy gain per branch can be obtained by an integral over momentum,
\begin{eqnarray}
\frac{L}{2\pi} \int_{0}^{k_F} dq \; \delta \epsilon_{q} \approx \frac{k_F L}{2 \pi} \frac{\Delta_{b}^2}{\hbar v_0 k_F} \ln \left| \frac{2\hbar v_0 k_F}{\Delta_{b}} \right| ,
\label{Eq:Peierls}
\end{eqnarray}
where we have kept the leading order in $\Delta_{b}/(\hbar v_0 k_F)$. Consequently, the bulk gap opening results in a Peierls energy gain. 
It is therefore energetically favorable that the system evolves from the crossed sliding Luttinger liquid to the \moire correlated state to open a gap at sufficiently low temperature. 

In consequence, whereas the Hamiltonian may allow for the formation of both chirality states, we expect that the system undergoes spontaneous symmetry breaking. At the transition, an infinitesimal perturbation will select one of the chirality states within a domain. This situation is analogous to the Ising ferromagnetic phase. While the ground state can have positive or negative magnetization, only one of them will be stabilized within a magnetic domain as the temperature is lowered through the transition from the paramagnetic to the ferromagnetic phase. As a result, we expect the formation of chirality domains in the \moire correlated state. 
While the domain walls cost energy, it is generally anticipated that domains can form, contributing to an increase in entropy and consequently a decrease in free energy at finite temperatures.  
Given that a chirality domain can carry finite orbital magnetization and couple to magnetic moments, we postulate that local magnetic moments (potentially induced by disorder) can initiate the formation of these domains. 
Remarkably, in Refs.~\citesupp{Sharpe:2019_S,Serlin:2020_S}, the stabilization of a definite chirality state in micrometer-sized systems requires a finite magnetic field to train domains in the samples.

It is worth mentioning that an alternative scenario is possible where the perturbation in Eq.~\eqref{Eq:Hab} realizes a self-dual sine-Gordon model~\citesupp{Lecheminant:2002_S}. While this leads to an intriguing possibility for stabilizing parafermion modes without superconductivity~\citesupp{Ronetti:2021_S}, the detailed analysis of the self-dual sine-Gordon model is beyond the scope of this work.

\subsection{IV. Experimental features for the edge modes}
In this section we discuss the experimental features for the edge modes when the system is in one of the \moire correlated states with a gapped bulk and gapless edges. 
For simplicity we consider here Eq.~\eqref{Eq:O_um4} with $N_{\uparrow}=N_{\downarrow}=M+1$, $M_{\uparrow}=M_{\downarrow}=M$, $n=1$, and a single domain in the system. 
This choice allows us to focus on a simpler case in which a single mode appears at one edge and to construct the effective edge theory from the commutator, given in Eq.~\eqref{Eq:commutator} in the main text~\citesupp{Kane:1996_S,Fisher:1997_S,Chang:2003_S}, 
\begin{eqnarray}
\frac{S_{\rm edge}}{\hbar} &=& \int \frac{ dx d \tau }{ 4\pi f}  \;  \Big[ -i \partial_x 
\phi \partial_\tau \phi + v_{\rm edge} \big( \partial_x  \phi \big)^2 \Big],
\hspace{5pt}
\end{eqnarray}
where $ \phi$ is the chiral boson field satisfying $ \big[ \phi (x), \phi (x') \big] = i  \pi f \, {\rm sign} (x-x')$
 at the edge and $v_{\rm edge}$ its velocity.

The local density of states is given by~\citesupp{Balents:1999_S,Fisher:1997_S} 
\begin{eqnarray}
\rho (\epsilon) &=& \frac{1}{\pi} {\rm Re}
 \left[ \int_0^{\infty} dt \; e^{i \epsilon t/\hbar} \Big\langle  \psi_{\rm e}(t) \psi_{\rm e}^{\dagger} (0)  \Big\rangle \right] ,  
 \end{eqnarray}
where $\psi_{\rm e} = e^{i \phi  /f} $ denotes the excitation operator with a unit charge. Computing the average in $\langle \cdots \rangle$ with respect to the action $S_{\rm edge}$ and performing the time integral, one gets the local density of states, which follows the universal scaling curve Eq.~\eqref{Eq:DOS} given in the main text and reduces to a power law $\rho (\epsilon) \propto |\epsilon|^{1/f-1} $ as $T \to 0$. 
The scaling behavior in the spectroscopic features can be verified through scanning tunneling spectroscopy, as earlier studies on carbon nanotubes~\citesupp{Bockrath:1999_S}.

Next, we discuss the transport features of the edge modes in the settings illustrated in Fig.~\ref{Fig:QPC} in the main text. 
The setting in Fig.~\ref{Fig:QPC}(a) allows for an interedge tunneling process, which can be described as
\begin{eqnarray}
S_{\rm t} &=& t_0 \int d \tau  \; e^{i ( \phi_1 - \phi_2 ) /f} ,
\end{eqnarray}
with the tunnel amplitude $t_0$. In the above, we include a subscript in the $\phi$ field to label the two edges. 
One can derive the RG flow equation for the tunnel amplitude,
\begin{eqnarray}
\frac{ d \tilde{t}_0 }{d l} &=& \Big(1 - \frac{1}{f} \Big) \tilde{t}_0 ,
\end{eqnarray}
where $\tilde{t}_0 = t_0 / \Delta_{\rm a}$ is the dimensionless coupling with the high-energy cutoff $\Delta_{\rm a} =  \hbar v_{\rm edge} / a$ and the dimensionless length scale $l$.
By integrating the RG flow equation up to the scale $l^{*}$ corresponding to the high-bias ($V$) or high-temperature ($T$) regime, one gets the renormalized coupling and differential tunneling conductance,
\begin{eqnarray}
\frac{ d I_{\rm t} }{d V} & \propto & \tilde{t}_0^2 (l^{*}) \propto {\rm max} (eV, k_BT)^{ \frac{2}{f} -2}.
\end{eqnarray}
Alternatively, one can directly compute the current through a tunneling barrier separating two Luttinger liquids~\citesupp{Hsu:2019_S} and obtain a universal scaling formula for general $V$ and $T$ given in Eq.~\eqref{Eq:I-V} in the main text. 

On the other hand, the QPC setting in Fig.~\ref{Fig:QPC}(b) leads to an interedge backscattering process,
\begin{eqnarray}
S_{\rm b} &=& v_{\rm b} \int d \tau  \; e^{i ( \phi_1 - \phi_2 ) }.
\end{eqnarray}
The RG flow equation for the backscattering strength $v_{\rm b}$ is  
\begin{eqnarray}
\frac{ d \tilde{v}_{\rm b} }{d l} &=& \big(1 - f \big) \tilde{v}_{\rm b},
\end{eqnarray}
with $\tilde{v}_{\rm b} = v_{\rm b} / \Delta_{\rm a}$.
It leads to a backscattering current in the opposite edge and therefore a correction $ \delta G<0$ in the edge (differential) conductance with the magnitude, 
\begin{eqnarray}
| \delta G|  & \propto &  \tilde{v}_{\rm b}^2 (l^{*}) \propto {\rm max} (eV, k_BT)^{2f -2}.
\end{eqnarray}

Crucially, as long as the operator $O_{\rm iv}$ is RG relevant, the scaling exponents do not depend on the details of the fixed-point Hamiltonian $H_0 + H_{\rm fs}$, a feature similar to edge transport of the fractional quantum Hall states~\citesupp{Kane:1996_S,Fisher:1997_S,Chang:2003_S}; we therefore do not explicitly give the form of $H_0 + H_{\rm fs}$ in the main text. As a note, for a general set of $(N_{\uparrow},N_{\downarrow},M_{\uparrow},M_{\downarrow})$ and $n$, there can be multiple gapless modes at an edge, and the forward scattering between the edge modes can lead to nonuniversal exponents~\citesupp{Kane:1994_S,Kane:1995_S}.

\subsection{V. Unperturbed Hamiltonian}
In this section we discuss a specific model describing the crossed sliding (Tomonaga-)Luttinger Liquid fixed point. 
In the literature~\citesupp{Emery:2000_S,Vishwanath:2001_S,Mukhopadhyay:2001a_S,Mukhopadhyay:2001b_S}, the following model was adopted, 
\begin{subequations}
\label{Eq:H0}
\begin{eqnarray}
H_0 + H_{\rm fs} &=& \sum_{j=1}^3  H_{c}^{(j)} + \sum_{j=1}^3  H_{s}^{(j)} ,  \nonumber  \\
H_{c}^{(j)} &=& \sum_{m m'} \int \frac{\hbar dx}{2\pi} \left[ V_{\phi,m m'}^{j} ( \partial_x \phi_{c m}^j ) (\partial_x \phi_{c m'}^j  )
\right. \nonumber \\
 && \hspace{52pt} \left. 
+ V_{\theta,m m'}^{j} ( \partial_x \theta_{c m}^j ) (\partial_x \theta_{c m'}^j ) \right],  \nonumber \\
H_{s}^{(j)} &=& \sum_{m} \int \frac{\hbar dx}{2\pi} \left[ \frac{u_s}{K_s} (\partial_x \phi_{s m}^j  )^2 + u_s K_s (\partial_x \theta_{s m}^j )^2 \right],  \nonumber \\
\end{eqnarray}
which has been modified for the TBG problem~\citesupp{Chen:2020_S}. 
For each array labeled by $j$, we separate the charge $H_{c}^{(j)} $ and spin $H_{s}^{(j)}$ sectors. 
The spin sector is characterized by the velocity $u_s$ and the interaction parameter $K_s$, set to be identical for all the wires. 
The charge sector contains the intrawire and interwire terms of the density-density $V_{\phi}^{j}$ and current-current $V_{\theta}^{j}$ interactions and can be Fourier transformed, 
\begin{eqnarray}
H_{c}^{(j)} &=& \frac{1}{\Omega_{\perp}} \sum_{ q_{\perp} } \int \frac{\hbar dx}{2\pi} \left[ \frac{u_c^{j} (q_{\perp})}{K_c^{j} (q_{\perp})} \Big|\partial_x \phi_c^{j}( q_{\perp} )\Big|^2 \right. \nonumber \\
&& \hspace{64pt} + \left. u_c^{j} (q_{\perp}) K_c^{j} (q_{\perp}) \Big|\partial_x \theta_c^{j}(q_{\perp})\Big|^2 \right], \nonumber \\
\label{Eq:H0c}
\end{eqnarray}
\end{subequations}
with the momentum $q_{\perp}$ in the perpendicular direction to the wire (that is, $\perp x$) and $\Omega_{\perp} = N_{\perp} d$. Here, the velocity $u_{c}^{j}$ and the interaction function $K_{c}^{j}$ (a generalization of the interaction parameter) are functions of $q_{\perp}$,
\begin{subequations}
\begin{eqnarray}
u_c^{j} (q_{\perp}) &\equiv& \sqrt{V_{\phi}^{j} (q_{\perp}) V_{\theta}^{j} (q_{\perp})}, \\
K_c^{j} (q_{\perp}) &\equiv& \sqrt{V_{\theta}^{j} (q_{\perp}) / V_{\phi}^{j} (q_{\perp})}.
\end{eqnarray}
\end{subequations}
For a system with the periodic boundary condition perpendicular to the wire (that is, parallel to $q_{\perp}$), it would be natural to express $V_{\phi}^{j}$ and $V_{\theta}^{j}$ as periodic functions of $q_{\perp}$. 
Nonetheless,
existing works took (the inverse of) the interaction function as a Fourier series of $q_{\perp}$~\citesupp{Emery:2000_S,Vishwanath:2001_S,Mukhopadhyay:2001a_S,Mukhopadhyay:2001b_S,Chen:2020_S},
\begin{eqnarray}
\frac{1}{K_c(q_{\perp})} &=&  \sum_{m=0}^{\infty} \kappa_{m} \cos (m q_{\perp} d),
\label{Eq:Kc_series} 
\end{eqnarray}
identical for all $j$ here. 
This choice makes it possible to express the scaling dimensions using the following dimensionless parameters, 
\begin{subequations}
\label{Eq:dimensionless}
\begin{eqnarray}
\Delta_{\phi n} &=& \int_{-\pi}^{\pi} \frac{d (q_{\perp} d)}{2\pi} K_c(q_{\perp}) \cos (n q_{\perp} d) ,\\
\Delta_{\theta n} &=& \int_{-\pi}^{\pi} \frac{d (q_{\perp} d)}{2\pi}  \frac{\cos (n q_{\perp} d)} {K_c( q_{\perp} )}.
\end{eqnarray}
\end{subequations}
We note that neglecting the marginally relevant interwire forward scattering terms amounts to truncate Eq.~\eqref{Eq:Kc_series} at $m=0$. 
In Ref.~\citesupp{Emery:2000_S}, the series was truncated at $m=1$.
In Refs.~\citesupp{Chen:2020_S,Vishwanath:2001_S,Mukhopadhyay:2001a_S,Mukhopadhyay:2001b_S}, the authors kept terms up to $m=2$, leading to
\begin{eqnarray}
\frac{1}{K_c( q_{\perp} )} &=& \frac{1}{K_{c0}} \big[ 1 + \lambda_1  \cos (q_{\perp} d) + \lambda_2  \cos (2 q_{\perp} d)  \big]. \hspace{20pt}
\label{Eq:Kc_explicit}
\end{eqnarray}

As discussed here, the choice of Eqs.~\eqref{Eq:H0}--\eqref{Eq:Kc_series} is not general. Therefore, instead of limiting our discussions to a particular model, we keep a general form for $H_0 + H_{\rm fs}$ in the main text. 
We note that, with this specific model, 
one can obtain a power-law $T$ dependence of the resistivity, as predicted in earlier works on similar models~\citesupp{Emery:2000_S,Mukhopadhyay:2001a_S,Mukhopadhyay:2001b_S}.

For the specific model listed in Eqs.~\eqref{Eq:H0}--\eqref{Eq:Kc_series}, we compute the scaling dimension of the general scattering operator $O_{\{s_{\ell p\sigma}\}} $ given in Eq.~\eqref{Eq:O_general} in the main text, 
\begin{eqnarray}
\Delta_{\{s_{\ell p\sigma}\}} &=& \frac{1}{8} \sum_{p,p'} \big[ S_{p,c} S_{p',c} \Delta_{\phi (p-p')} + \bar{S}_{p,c} \bar{S}_{p',c} \Delta_{\theta (p-p')} \big]  \nonumber \\
&& + \frac{1}{8} \sum_{p} \big[ S_{p,s}^2 K_{s} + \frac{\bar{S}_{p,s}^2}{ K_{s} } \big],
\end{eqnarray} 
with the coefficients
\begin{subequations}
\begin{eqnarray}
 S_{p,c} &=& s_{Lp\uparrow} - s_{Rp\uparrow}  + s_{Lp\downarrow} - s_{Rp\downarrow}, \\
 \bar{S}_{p,c} &=& s_{Lp\uparrow} + s_{Rp\uparrow}  + s_{Lp\downarrow} + s_{Rp\downarrow}, \\
 S_{p,s} &=& s_{Lp\uparrow} - s_{Rp\uparrow}  - s_{Lp\downarrow} + s_{Rp\downarrow}, \\
 \bar{S}_{p,s} &=& s_{Lp\uparrow} + s_{Rp\uparrow}  - s_{Lp\downarrow} - s_{Rp\downarrow}.
\end{eqnarray}
\end{subequations}
When the interaction function $K_c ( q_{\perp} )$ is specified, such as the one in Eq.~\eqref{Eq:Kc_explicit}, the integrals in Eq.~\eqref{Eq:dimensionless} can be computed directly. 

For scattering processes within an array, the RG relevance for the corresponding operators is determined by
\begin{eqnarray}
\Delta_{\{s_{\ell p\sigma}\}} &<& 2, 
\end{eqnarray}
where the right-hand side comes from the temporal and spatial integrals in the action~\citesupp{Giamarchi:2003_S}. When the operator $O_{\{s_{\ell p\sigma}\}} $ is RG relevant, the crossed sliding Luttinger liquid is unstable against the perturbation, leading to an electronic state characterized by $O_{\{s_{\ell p\sigma}\}} $.
For interarray scattering processes, which take place at wire intersections, on the other hand, the RG relevance condition becomes $\Delta_{\{s_{\ell p\sigma}\}} <1$, owing to the lack of the spatial integral in the action. 
Therefore, the RG relevance condition is stricter than those within an array; 
we therefore examine the scatterings within an array in more detail throughout the main text.

\bibliographysupp{supp}

\end{document}